\documentclass[conference]{IEEEtran}
\renewcommand\IEEEkeywordsname{Index Terms}
\usepackage{graphicx}
\usepackage{subcaption}
\usepackage{tikz,xparse}
\usetikzlibrary{dsp,chains}
\usetikzlibrary{matrix}
\usepackage{mathptmx}
\usepackage{verbatim}
\usepackage{calc}
\usepackage{ifthen}
\usepackage{xifthen}
\usepackage{cancel}
\usepackage{bm}
\usepackage{verbatim}
\usepackage{multirow}
\usepackage{cite}

\usepackage[nolist]{acronym} 
\usepackage{pgfplots}
\usetikzlibrary{arrows,shapes,graphs,graphs.standard,quotes}
\usetikzlibrary{matrix}
\pgfplotsset{compat=newest}

\usepackage[hyphens]{url}

\usepackage[bookmarks=false]{hyperref}
\usepackage{units}
\usepackage{amsmath, amsbsy, amssymb, latexsym }
\hypersetup{bookmarksdepth=-2}
\usepackage{comment}
\usepackage[utf8]{inputenc}
\usepackage{xcolor}
\usepackage{enumitem}
\usepackage{algorithm}
\usepackage{algpseudocode}

\usepackage{etoolbox}

\tikzset{>=latex}

\algnewcommand\algorithmicforeach{\textbf{for each}}
\algdef{S}[FOR]{ForEach}[1]{\algorithmicforeach\ #1\ \algorithmicdo}

\algnewcommand\algorithmicswitch{\textbf{switch}}
\algnewcommand\algorithmiccase{\textbf{case}}
\algnewcommand\algorithmicassert{\texttt{assert}}
\algnewcommand\Assert[1]{\State \algorithmicassert(#1)}%
\algdef{SE}[SWITCH]{Switch}{EndSwitch}[1]{\algorithmicswitch\ #1\ \algorithmicdo}{\algorithmicend\ \algorithmicswitch}%
\algdef{SE}[CASE]{Case}{EndCase}[1]{\algorithmiccase\ #1}{\algorithmicend\ \algorithmiccase}%
\algtext*{EndCase}%

\definecolor{mittelblau}{RGB}{0, 126, 198}
\definecolor{violettblau}{cmyk}{0.9, 0.6, 0, 0}
\definecolor{rot}{RGB}{238, 28 35}
\definecolor{apfelgruen}{RGB}{140, 198, 62}
\definecolor{gelb}{RGB}{255, 229, 0}
\definecolor{orange}{RGB}{244, 111, 33}
\definecolor{pink}{RGB}{237, 0, 140}
\definecolor{lila}{RGB}{128, 10, 145}
\definecolor{hellgrau}{RGB}{224, 224, 224}
\definecolor{mittelgrau}{RGB}{128, 128, 128}
\definecolor{dunkelgrau}{RGB}{80,80,80}
\definecolor{anthrazit}{RGB}{19, 31, 31}
\definecolor{darkgreen}{RGB}{34,139,34}
\colorlet{Mycolor1}{green!10!orange!90!}
\tikzset{
       vnd/.style={
        shape=circle,
        fill=black,
        draw,
        inner sep=0pt,
        minimum size=0.2cm},
        cnd/.style={
        shape=rectangle,
        fill=white,
        draw,
        minimum width=0.05mm,
        minimum height = 0.05mm}, 
         vndR/.style={
        shape=circle,
        fill=red,
        draw,
        inner sep=0pt,
        minimum size=0.2cm},
        cndR/.style={
        shape=rectangle,
        fill=white,
        draw=red,
        minimum width=0.05mm,
        minimum height = 0.05mm}
}

\IEEEoverridecommandlockouts

\renewcommand{\vec}[1]{\mathbf{#1}}

\newcommand{\rv}{\vec{r}}

\newcommand{\uv}{\vec{u}}

\newcommand{\xv}{\vec{x}}
\newcommand{\yv}{\vec{y}}

\newcommand{\Gm}{\vec{G}}

\begin{document}

\begin{NoHyper}
\title{Genetic Algorithm-based Polar Code Construction for the AWGN Channel\vspace{-0.5cm}}

\author{\IEEEauthorblockN{Ahmed Elkelesh, Moustafa Ebada, Sebastian Cammerer and Stephan ten Brink} \thanks{This work has been supported by DFG, Germany, under grant BR 3205/5-1. The authors would like to thank Alexander Vardy and Ido Tal for providing their set of frozen/non-frozen bit positions as a baseline for comparison.}
	\IEEEauthorblockA{
		Institute of Telecommunications, Pfaffenwaldring 47, University of  Stuttgart, 70569 Stuttgart, Germany 
		\\\{elkelesh,ebada,cammerer,tenbrink\}@inue.uni-stuttgart.de
	}
}

\makeatletter
\patchcmd{\@maketitle}
{\addvspace{0.5\baselineskip}\egroup}
{\addvspace{-0.6\baselineskip}\egroup}
{}
{}
\makeatother

\maketitle

\begin{acronym}
 \acro{ECC}{error-correcting code}
 \acro{HDD}{hard decision decoding}
 \acro{SDD}{soft decision decoding}
 \acro{ML}{maximum likelihood}
 \acro{GPU}{graphical processing unit}
 \acro{BP}{belief propagation}
 \acro{BPL}{belief propagation list}
 \acro{LDPC}{low-density parity-check}
  \acro{HDPC}{high density parity check}
 \acro{BER}{bit error rate}
 \acro{SNR}{signal-to-noise-ratio}
 \acro{BPSK}{binary phase shift keying}
 \acro{AWGN}{additive white Gaussian noise}
 \acro{MSE}{mean squared error}
 \acro{LLR}{Log-likelihood ratio}
 \acro{MAP}{maximum a posteriori}
 \acro{NE}{normalized error}
 \acro{BLER}{block error rate}
 \acro{PE}{processing elements}
 \acro{SCL}{successive cancellation list}
 \acro{SC}{successive cancellation}
 \acro{BI-DMC}{Binary Input Discrete Memoryless Channel}
 \acro{CRC}{cyclic redundancy check}
 \acro{BEC}{Binary Erasure Channel}
 \acro{BSC}{Binary Symmetric Channel}
 \acro{BCH}{Bose-Chaudhuri-Hocquenghem}
 \acro{RM}{Reed--Muller}
 \acro{RS}{Reed-Solomon}
  \acro{SISO}{soft-in/soft-out}
\acro{PSCL}{partitioned successive cancellation list}
  \acro{3GPP}{3rd Generation Partnership Project }
  \acro{eMBB}{enhanced Mobile Broadband}
      \acro{CN}{check nodes}
      \acro{PC}{parity-check}
      \acro{GenAlg}{Genetic Algorithm}
\acro{AI}{Artificial Intelligence}
\acro{MC}{Monte Carlo}
\acro{CSI}{Channel State Information}
\acro{PSCL}{partitioned successive cancellation list}
\end{acronym}

\begin{abstract}
	
We propose a new polar code construction framework (i.e., selecting the frozen bit positions) for the \ac{AWGN} channel, tailored to a given decoding algorithm, rather than based on the (not necessarily optimal) assumption of \ac{SC} decoding.
The proposed framework is based on the \ac{GenAlg}, where populations (i.e., collections) of information sets evolve successively via evolutionary transformations based on their individual error-rate performance. These populations converge towards an information set that fits the decoding behavior.
Using our proposed algorithm, we construct a polar code of length $2048$ with code rate $0.5$, without the CRC-aid, tailored to \emph{plain} \ac{SCL} decoding, achieving the same error-rate performance as the CRC-aided SCL decoding, and leading to a coding gain of $\unit[1]{dB}$ at BER of $10^{-6}$. Further, a \ac{BP}-tailored polar code approaches the \ac{SCL} error-rate performance without any modifications in the decoding algorithm itself.
\end{abstract}
\acresetall
\vspace{-0.49cm}
\section{Introduction}
\vspace{-0.2cm}

Polar codes \cite{ArikanMain} are the first family of codes proven to be capacity achieving for any \ac{BI-DMC} and with an explicit construction method under low complexity \ac{SC} decoding. However, for finite lengths, both the \ac{SC} decoder and the polar code itself (i.e., its \ac{ML} performance) are shown to be sub-optimal when compared to other state-of-the-art coding schemes such as \ac{LDPC} codes. Later, Tal and Vardy introduced a \ac{SCL} decoder \cite{talvardyList} enabling a decoding performance close to the \ac{ML} bound for sufficiently large list sizes. The concatenation with an additional high-rate \ac{CRC} code \cite{talvardyList} or \ac{PC} code \cite{PCC} further improves the code performance itself, as it increases the minimum distance and, thus, improves the weight spectrum of the code. This simple concatenation renders polar codes into a powerful coding scheme. Later, an extension of polar codes, namely Polar Subcodes, were proposed in \cite{subCodes}, outperforming the above-mentioned code constructions. Polar codes were recently selected by the 3GPP group as the channel codes for the upcoming \emph{5th generation mobile communication standard (5G)} uplink/downlink control channel \cite{Huawei}.

Although several other decoding schemes such as the iterative \ac{BP} decoding \cite{ArikanBP} and the iterative \ac{BPL} decoder \cite{elkelesh2018belief} exist, their error-rate performance is currently not competitive with the CRC-aided SCL decoding. Therefore, a good polar code design tailored to an explicit decoder may change this situation as it promises either an improved error-rate performance or savings in terms of decoding complexity for a specific type of decoder. 
In this work, we consider the decoding algorithm throughout the code construction phase instead of constructing the code based on the typical assumption of \ac{SC} decoding. 
An intuitive example why a design for \ac{SC} is sub-optimal for other decoding schemes can be given by considering the girth of the graph under \ac{BP} decoding, which strongly depends on the frozen positions of the code. Thus, freezing additional nodes can even result in a degraded decoding performance under \ac{BP} decoding although the \ac{ML} performance indeed gets better (see Fig.~9 in \cite{ISWCS_Error_Floor}).

In a strict sense, one may argue that polar codes are inherently connected with \ac{SC} decoding and only a design based on the concept of channel polarization results in a \emph{true} ``polar'' code. However, from a more practical point of view, we seek to find the most efficient coding scheme for finite length constraints and, with slight abuse of notation, we regard the proposed codes as polar codes. Thus, a design method that considers the decoder and improves the overall performance is an important step for future polar code applications.

Polar code construction (or design), throughout this paper, refers to selecting an appropriate frozen bit position pattern. Different polar code construction algorithms exist assuming SC decoding. However, an explicit design tailored to \ac{SCL} or \ac{BP} decoding turns out to be cumbersome due to the many dependencies in the decoding graph and the high dimensionality of the optimization problem. In this work, we propose a new framework for polar code design matched to a specific decoding algorithm embedded in the well-understood \ac{GenAlg} context. As a result, the optimization algorithm works on a specific error-rate simulation setup, i.e., it inherently takes into account the actual decoder and channel. This renders the \ac{GenAlg}-based polar code optimization into a solid and powerful design method. To the best of our knowledge, the resulting polar codes in this work outperform any known design method for \emph{plain} polar codes under \ac{SCL} decoding \emph{without} the aid of an additional \ac{CRC} (i.e., CRC-aided SCL performance could be achieved \emph{without} the aid of a \ac{CRC}). Additionally, the \ac{BP} decoder of the \emph{proposed} code achieves (and slightly outperforms) the \ac{SCL} decoding performance of \emph{conventional} codes without any decoder modifications. 
The interested reader can refer to \cite{GenAlg_Journal} for an extended version.

\section{Polar codes} \label{sec:polarcodes}
\vspace{-0.2cm}
Polar codes \cite{ArikanMain} are based on the concept of channel polarization, in which $N=2^n$ identical copies of a channel $W$ are combined and $N$ synthesized bit-channels are generated. These synthesized bit-channels show a polarization behavior, in the sense that some bit-channels are purely noiseless and the rest are completely noisy.
A recursive channel combination provides the polarization matrix 
\begin{align*} 
\mathbf{G}_N = \mathbf{B}_N \cdot \mathbf{F}^{\otimes n}, \qquad \mathbf{F} = \left[ \begin{array}{ll} 1 & 0 \\ 1 & 1 \end{array}\right] 
\end{align*}
where $\mathbf{B}_N$ is a bit-reversal permutation matrix and $\mathbf{F}^{\otimes n}$ denotes the $n$-th Kronecker power of $\mathbf{F}$. 
The polar codewords $\xv$ are given by $\xv = \uv \cdot \Gm_N$, where $\mathbf{u}$ contains $k$ information bits and $N-k$ frozen bits, w.l.o.g. we set the frozen positions to ``0''. 
The information set $\mathbb{A}$ contains the $k$ most reliable positions of $\uv$ in which the $k$ information bits are transmitted and $\bar{\mathbb{A}}$ denotes the frozen positions (i.e., the complementary set to $\mathbb{A}$). The \emph{conventional} generator matrix, denoted by $\Gm$, is constructed as the rows $\left\{\rv_i\right\}$ of $\Gm_N$ with $i\in\mathbb{A}$.
The task of the polar code construction, in its original form, is to find the information set $\mathbb{A}$ which maximizes the code performance (under \ac{SC} decoding) for a specific channel condition. More details on the problem of polar code construction is provided in Section \ref{sec:con}.

Throughout this work, we use non-systematic polar encoding. However, it is straightforward to use the \ac{GenAlg} to construct systematic polar codes. In this work, a polar code with codeword length $N$ and $k$ information bits is denoted by $\mathcal{P} \left(N,k\right)$, i.e., the information set has the cardinality $|\mathbb{A}|=k$ and the code rate $R_c = k/N$. In the following, we revise the basic polar decoding algorithms.

\textit{\ac{SC} decoding} \cite{ArikanMain} is the first polar decoder, where all information bits $\hat{u}_i$ are sequentially hard-decided based on the previously estimated bits $\left\{\hat{u}_1,\dots,\hat{u}_{i-1}\right\}$ and the channel information $\yv$, where $i\in\left\{1,\dots,N\right\}$. 

\textit{\ac{SCL} decoding} \cite{talvardyList} denoted by \ac{SCL} $\left(L\right)$, utilizes a list of $L$ most likely paths during \ac{SC} decoding; at every decision the decoder branches into two paths ($\hat{u}_i=0$ and $\hat{u}_i=1$) instead of the hard decision in the \ac{SC} decoder. To limit the exponential growth of complexity, only the $L$ most reliable paths are kept sorted in the list according to a specific path metric. 

\textit{CRC-aided SCL decoding} \cite{talvardyList} denoted by \ac{SCL}+\ac{CRC}-$r$ $\left(L\right)$, where an additional high-rate \ac{CRC} of $r$ bits is concatenated to the polar code, to help in selecting the final codeword from the $L$ surviving candidates, yielding significant performance gains in competing with the state-of-the-art error correcting codes.

\textit{\ac{BP} decoding} \cite{ArikanBP} denoted by \ac{BP} $\left(N_{it,max}\right)$ is an iterative message passing decoder based on the \emph{encoding} graph of the polar code. \ac{LLR} messages are iteratively passed along the encoding graph until reaching a maximum number of iterations $\left(N_{it,max}\right)$ or meeting an early stopping condition. The error-rate performance of polar codes under \ac{BP} decoding is typically not competitive with \ac{SCL} decoding. In this work, we use the \emph{$\mathbf{G}$-matrix-based} early stopping condition, where decoding terminates when $\mathbf{\hat{x}} = \mathbf{\hat{u}}\cdot\mathbf{G}_N$ \cite{earlyStop}.

\vspace{-0.1cm}
\section{Polar Code Construction} \label{sec:con}
The polar code construction phase is about deciding the most reliable $k$ bit positions that are set as the information bit positions, while the remaining $N-k$ bit positions are set as frozen bit positions. Thus, ranking the bit-channels according to their reliabilities is of major significance in the polar code construction phase. 
The information set $\mathbb{A}$ is the outcome from the polar code construction phase specifying the indices of the information bit positions. A corresponding logical $\mathbf{A}$-vector can be used such that $\mathbf{A} = \left[ a_1,a_2,\dots,a_N \right]$, where $a_i \in \left \{ 0,1 \right \}$ and $1 \le i \le N$. Bit position $i$ is frozen if $a_i = 0$, while bit position $j$ is non-frozen (i.e., can be used for information transmission) if $a_j = 1$. For instance consider the $\mathcal{P} \left(8,4\right)$-code, the information set  $\mathbb{A} = \left \{ 4,6,7,8 \right \}$ can thus be represented by the logical vector $\mathbf{A} = \left[ 0\,0\,0\,1\,0\,1\,1\,1 \right]$.

The code construction can be considered as an optimization problem, where the objective is to find the optimal set of $k$ good positions in a set of indices $\left\{1,\dots,N\right\}$, as shown in (\ref{PolarCodeConstruction}). 
\vspace{-0.55cm}

\begin{equation}
	\begin{aligned}
		& \mathbf{A}_{\text{opt}} =\arg \underset{\mathbf{A}}{\text{      min}}
		& & \mathrm{BER}(\mathbf{A}) \text{ or } \mathrm{BLER}(\mathbf{A})\\
		& \text{subject to}
		&& \left( \sum_{i=1}^{N} a_i \right) = k, \\
		&&& a_i \in \left \{ 0,1 \right \}. \\
	\end{aligned}
	\label{PolarCodeConstruction}
\vspace{-0.05cm}
\end{equation}
where $\mathbf{A} = \left[ a_1,a_2,\dots,a_N \right]$, $R_c = k / N$ and $i = 1,2,\dots,N$.

The information set $\mathbb{A}$ (or, equivalently, the $\mathbf{A}$-vector) is channel dependent, meaning that it depends on the respective channel parameter (e.g., design SNR for \ac{AWGN} channel, or design $\epsilon$ for \ac{BEC}). Thus, polar codes are non-universal.
The minimum distance of polar codes $d_{min}$ depends on the polar code construction (i.e., $\mathbb{A}$). Polar codes $d_{min}$ is equal to the minimum weight of the rows $\left\{\mathbf{r}_i\right\}$ in the $\mathbf{G}_N$-matrix with indices in $\mathbb{A}$ (i.e., $i\in\mathbb{A}$) \cite[Lemma 3]{Urbanke_chCsC_BP}.

Choosing the best $k$ bit positions for information transmission is even more crucial for short length polar codes. This can be attributed to the fact that the bit-channels of the short length polar codes are not fully polarized, and the portion of semi-polarized bit-channels (which would be normally unfrozen) leads to high error-rates. Although efficient polar code construction only exists for the \ac{BEC} case \cite{ArikanMain}, many algorithms were devised for the \ac{AWGN} channel case. A survey on the effect of the design SNR and the effect of the specific polar code construction algorithm used on the error-rate performance of \ac{SC} decoding is presented in \cite{Vangala}.

Arıkan uses the symmetric capacity $I(W_i)$ or the Bhattacharyya parameter $Z(W_i)$ of the virtual channel $W_i$ to assess its reliability \cite{ArikanMain}. However, the Bhattacharyya parameters are preferred because of being connected with an explicit bound on the \ac{BLER} under \ac{SC} decoding.

In \cite{constructDE}, the polar code construction step is viewed as an instance of density evolution and, thus, a construction algorithm based on convolutions was proposed. A Gaussian approximation of the density evolution for polar code construction was proposed in \cite{constructGaussian}. 

In \cite{Urbanke_chCsC_BP}, picking the frozen bit positions according to the \ac{RM} rule was observed to enhance the error-rate performance significantly under \ac{MAP} decoding due to the fact that the \ac{RM} rule maximizes the minimum distance $d_{min}$. An \ac{RM} code can be viewed as a polar code with a different frozen/non-frozen bit selection strategy \cite{Urbanke_chCsC_BP}, where both codes are based on the same polarization matrix $\mathbf{G}_N$. However,  the $k$ information bit positions of  \ac{RM} codes are the positions corresponding to the $k$ row indices with the maximum weights in the $\mathbf{G}_N$-matrix. Consequently, the RM code construction phase is channel independent as it merely depends on the row weights of $\mathbf{G}_N$. A hybrid polar and RM code construction \cite{HybridTse} results in significant error-rate improvement gains under \ac{SCL} decoding without the \ac{CRC}-aid, by improving the minimum distance of the resultant code. This underlines the benefits of improving the minimum distance of the code and also the need of an improved construction algorithm tailored to the decoder. A similar family of codes was introduced in \cite{RMurbankePolar} providing significant error-rate performance gains under \ac{BP} and \ac{SCL} decoding.

In \cite{BP_LLR_Siegel}, an \ac{LLR}-based polar code construction is proposed, in which the \ac{LLR}s of the non-frozen bits are tracked during \ac{BP} decoding to identify weak information bit-channels, and then the information set $\mathbb{A}$ is modified by swapping these weak information bit-channels with strong frozen bit-channels. The resulting code shows an enhanced error-rate performance under both \ac{BP} and \ac{SCL} decoding due to the resultant reduction in the number of low weight codewords. Some work has been done to construct polar codes which are tailored to a specific decoder, e.g., polar code construction assuming \ac{SCL} decoding \cite{TUM_SCL_Construct} and \ac{BP} decoding \cite{MC_BP_2}, where a Monte-Carlo-based construction is proposed similar to \cite{ArikanMain}.

As the output alphabet size grows exponentially with the codelength, it is computationally of high complexity to precisely calculate $I(W_i)$ or $Z(W_i)$ per bit-channel. However, a quantization can be used to closely approximate them \cite{constructTalVardy}.
A recent discovery which reduces the complexity of the polar code design is the partial order for synthesized channels \cite{UPO}.
Later an heuristic closed-form algorithm called polarization weight (PW) \cite{PW} was proposed to determine the reliability of bit-channels based on their indices and a carefully chosen parameter $\beta$ \cite{BetaIngmard}, resulting in a significant complexity reduction in the polar code construction. 

It is important to keep in mind that polar codes that are constructed based on the mutual information or the Bhattacharyya parameters of the bit-channels, as proposed by Arıkan, are tailored to hard-output \ac{SC} decoders. Therefore, they are not necessarily optimum when using other decoders such as the soft-output \ac{BP} decoder \cite{Urbanke_chCsC_BP,RMurbankePolar,BP_LLR_Siegel} or the \ac{SCL} decoder \cite{RMurbankePolar,BP_LLR_Siegel,polarDesign5G}. To the best of our knowledge, no analytical polar code construction rule exists thus far which would be \emph{optimized} for \ac{BP} or \ac{SCL} decoding and, thus, the nature of the iterative or list decoding is not usually taken into account while designing polar codes. Therefore, the problem of taking the type of decoding into consideration while constructing the polar code is, thus far, an open problem. In this work, we propose a method which always converges to a ``good enough'' solution, i.e., leads to a better error-rate performance when compared to the state-of-the-art polar code construction techniques. 
\section{Genetic Algorithm-based polar code construction} \label{sec:genAlg}

\ac{GenAlg} was first introduced by Holland in 1975 \cite{GeneticsFirstpaper} as an efficient tool that helps in achieving (good) solutions for high-dimensional optimization problems which are computationally intractable. Beside finding (good) local minima, a well-parametrized \ac{GenAlg} is known for converging to these minima very quickly \cite{GAturbo}. Due to that merit, \ac{GenAlg} has attracted a lot of research in the \ac{AI} field leading to improved and adaptive variants of it. 

Furthermore, researchers from other fields (e.g., channel coding, signal processing) worked on adapting the \ac{GenAlg} to some of their specific problems that lacked enough theoretical basis for solving, leading to improvements over the existent theoretical methods of approaching the same problems. \ac{GenAlg} was applied in the channel coding field, e.g., during code construction \cite{GAMRD,GAturbo} and while decoding of linear block codes \cite{GAlinearBlockCodes}, \ac{LDPC} codes \cite{GAldpcdec} and convolutional codes \cite{GAconv}.

The \ac{GenAlg} is inspired by the natural evolution where \emph{populations} of \emph{offsprings}, resulting from \emph{parents}, keep \emph{evolving} and compete, while only the \emph{fittest} offsprings \emph{survive}. 
Therefore, the \ac{GenAlg} starts with an initial population of individuals and the fittest of them survive and give birth to new offsprings which represent the new population. The criterion of fitness is, therefore, critical for the \ac{GenAlg} so that the offsprings keep evolving towards the fittest. If the size of population and the number of evolution stages are sufficiently large, the \ac{GenAlg} converges to an ultimate (sub-) optimal solution.

The task of polar code construction can be viewed as an optimization problem (see (\ref{PolarCodeConstruction})) searching for the optimum information set $\mathbb{A}$ that has the minimum (possible) cost function. This optimization problem can be solved using \ac{GenAlg}. The \ac{BER} has been selected as the cost function throughout this work for optimization conducted on \ac{AWGN} channels. All presented results throughout this work are simulated on GPUs to accelerate our error-rate simulations \cite{cammerer_HybridGPU}. The whole setup is depicted in Fig. \ref{fig:block-diagram}. Next we briefly introduce the most important \ac{GenAlg} fundamentals.

\begin{figure}[t]	
	\centering
	\resizebox{0.9\columnwidth}{!}{
		\includegraphics{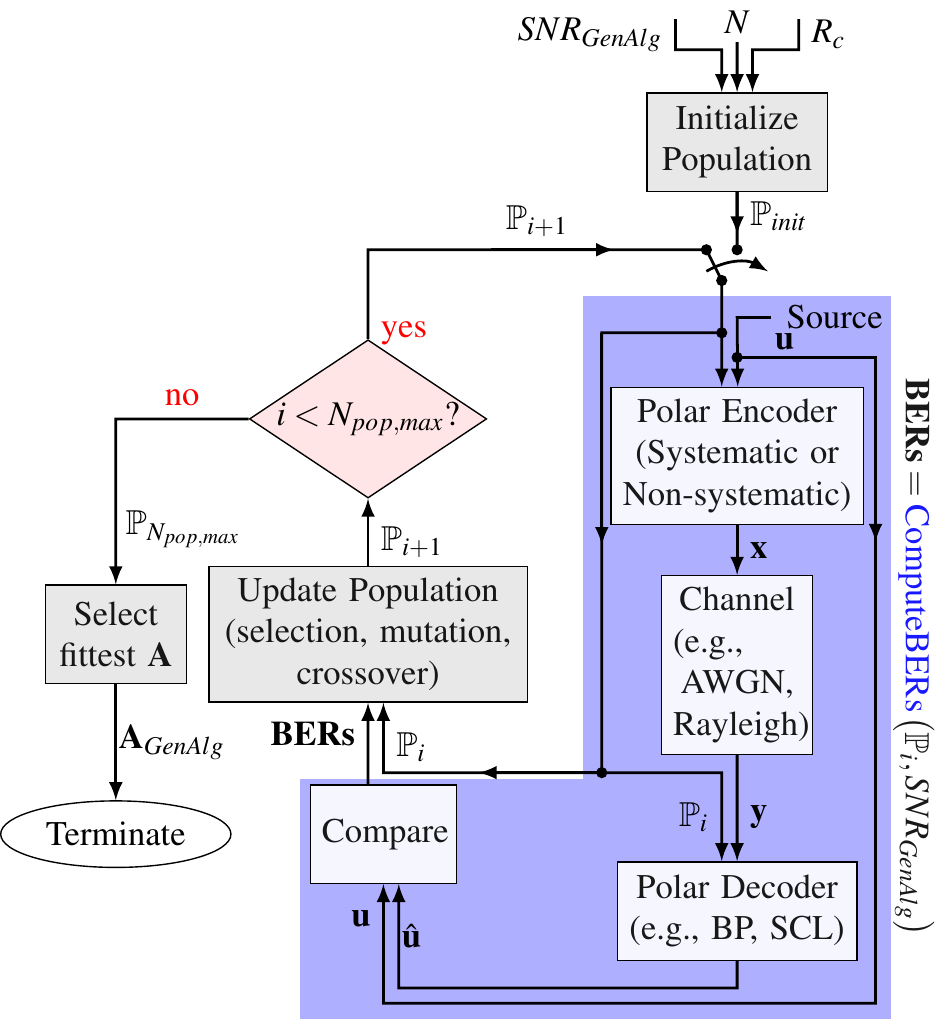}		
	} 
	\vspace{-0.1cm}
	\caption{\small An abstract view of the genetic algorithm (GenAlg)-based polar code construction.}
	\label{fig:block-diagram}  
	\vspace{-0.6cm}	
\end{figure}

\vspace{-0.25cm}

\subsection{Population} \vspace{-0.1cm}
In this work, the population $\mathbb{P}=\left\{\mathbf{A}_i\right\}$, for $i=1,\dots,S$, is a collection of $S$ candidate information sets represented by their binary $\mathbf{A}$-vectors, each with its own fitness value (e.g., \ac{BER} or \ac{BLER}), that represent the search space.
\vspace{-0.2cm}
\subsection{Initialization}
\vspace{-0.15cm}

As it turns out, \ac{GenAlg} converges faster, and probably to a better solution, if its population $\mathbb{P}_{init}$ is initialized with a sufficiently good collection of estimated $\mathbf{A}$-vectors. For that purpose, the population is initially filled with a collection of $\mathbf{A}$-vectors, all based on the Bhattacharyya construction \cite{ArikanMain} obtained for BECs with various erasure probabilities. Besides, we considered having the $\mathbf{A}$-vectors constructed according to \cite{HybridTse} among the initial population, in the \ac{SCL}-tailored polar code construction phase, which improved the acquired solution $\mathbf{A}_{GenAlg}$ remarkably.
\vspace{-0.2cm}
\subsection{Fitness function}
\vspace{-0.15cm}
The cost function chosen in this work is the error-rate at a user-defined design SNR $\left(SNR_{GenAlg}\right)$. The fitness function that decides the rank of each of the individual $\mathbf{A}$-vectors is, thus, selected to be the inverse of the error-rate. In other words, the $\mathbf{A}$ leading to the minimum error-rate is announced to be the optimum $\mathbf{A}_{GenAlg}$. Alternatively, one might consider using mutual information, \ac{LLR}-reliability $\sum_{i=1}^{N}\left|LLR_i\right|$ or any other reasonable metric as the fitness function.
\vspace{-0.3cm}
\subsection{Mutation}
\vspace{-0.1cm}
Mutation guarantees more diversity and acts against premature convergence at a certain bit position. It is, straightforwardly, a bit flip of a random position in the $\mathbf{A}$-vector representing a \emph{frozen-to-non-frozen} (or a \emph{non-frozen-to-frozen}) switch at that respective bit position. A mutation example is shown in Fig. \ref{fig:mutation}, where the offspring vector $\mathbf{Y}$ is the result of bit flipping the $2^{nd}$ bit of the parent vector $\mathbf{X}$. 
However, the polar code construction problem has the constraint that the number of non-frozen bit positions (i.e., number of ones) is equal to $k$ to maintain the code rate $R_c=k/N$. To restore $R_c$, one further mutation is applied to the resultant offspring vector $\mathbf{Y}$ yielding the vector $\mathbf{Z}$. For our specific problem, one mutation always occurs for each parent per evolutionary step.

\begin{figure}[t]
	\vspace{-0.0cm}
	\captionsetup[subfigure]{position=b}
	\centering
	
	\begin{subfigure}{0.425\columnwidth}
		\centering 
		\includegraphics{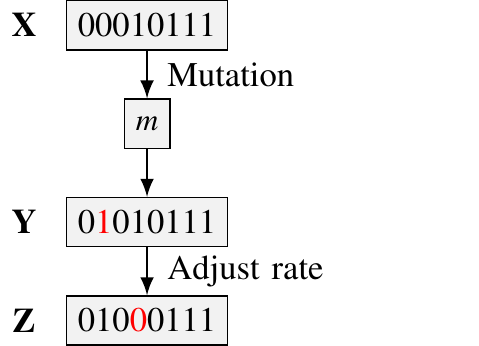}		
		\vspace{-0.3cm} \caption{ Mutation}
		\label{fig:mutation}
	\end{subfigure}   \hspace{-1cm}
	\begin{subfigure}{0.425\columnwidth}
		\centering
		\includegraphics{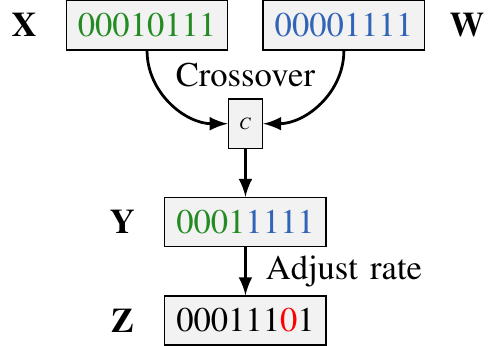}		
		\vspace{-0.3cm}\caption{Crossover}
		\label{fig:crossover}
	\end{subfigure}   
	
	\vspace{-0.1cm}
	\caption{\small Examples of \ac{GenAlg}'s evolutionary transformations in the polar code construction context. Inputs $\left(\mathbf{X} \ \text{and/or} \ \mathbf{W}\right)$ and output $\left(\mathbf{Z}\right)$ satisfy the constraints in (\ref{PolarCodeConstruction}).}
	\label{fig:block-diagram-mut-cross}
	\vspace{-0.6cm}
\end{figure}
\vspace{-0.3cm}
\subsection{Crossover}
\vspace{-0.1cm}
The crossover applied throughout this work is a single midpoint crossover (see Fig. \ref{fig:crossover}), where the $1^{st}$ half of the first parent vector $\mathbf{X}$ is combined with the $2^{nd}$ half of the second parent vector $\mathbf{W}$ to generate the vector $\mathbf{Y}$. 
This often leads to a change in the number of ones in the resulting vector $\mathbf{Y}$ (i.e., remember from (\ref{PolarCodeConstruction}) that the total number of ones in the $\mathbf{A}$-vector should be equal to $k$ and thus the code rate $R_c=k/N$). To restore $R_c$, sequential mutations are applied to the resultant vector until reaching the $k$ ones in the binary $\mathbf{A}$-vector ($\mathbf{Z}$ in Fig. \ref{fig:crossover}). Similar to mutation, one crossover always occurs for each pair of parents per evolutionary step. 

\subsection{Selection and population update}
In this context, by the term \emph{selection} we mean the way the new population is generated. We applied the following scheme in order to generate the new population:
\begin{itemize}
	\item The fittest $T$ $\mathbf{A}$-vectors are always pushed forward as members of the new population (i.e., self-offsprings). This ensures convergence to the (local) optimum and guarantees a monotonic behaviour of the cost function through evolving populations which facilitates observing the candidate solutions.
	\item Crossovers are applied between each pair of the  fittest $T$ $\mathbf{A}$-vectors, resulting in new ${T \choose 2}$ offsprings.
	\item Mutations are applied on the  fittest $T$ $\mathbf{A}$-vectors, resulting in new $T$ mutated offsprings.
\end{itemize}
Consequently, the size of the new population $S$ is 
\begin{align*}S=\underset{T \ fittest}{\underbrace{T}}+\underset{crossover}{\underbrace{{T\choose 2}}} + \underset{mutation}{\underbrace{T}} = \dfrac{T^2 +3T}{2} \label{Eq:S}\end{align*}
For all simulation results using the \ac{GenAlg} as discussed next, we set $S=20$ and $T=5$. We make the source code public and also provide the best polar code designs from this work online\footnote{\url{https://github.com/AhmedElkelesh/Genetic-Algorithm-based-Polar-Code-Construction}}.
\section{Simulation Results over \ac{AWGN} channel} \label{sec:resAWGN}
In this section, we show the results of designing polar codes using the \ac{GenAlg} method over the \ac{AWGN} channel. To be coherent with the results shown in \cite{talvardyList}, we use codes of length $N=2048$ and code rate $R_c=0.5$. 
\subsection{\ac{BP} decoder}
We use the \ac{GenAlg} to design polar codes tailored to \ac{BP} decoding over \ac{AWGN} channel. Fig. \ref{fig:BP_comp} shows a \ac{BER} comparison between a code constructed via \cite{constructTalVardy} at a design SNR $\left({E_b}/{N_0}\right) = \unit[2]{dB}$ and a code constructed using our proposed \ac{GenAlg} at $SNR_{GenAlg}$ $\left({E_b}/{N_0}\right) = \unit[2]{dB}$ under BP $\left(N_{it,max}=200\right)$ decoding. The \ac{GenAlg}-based construction yields a $\unit[0.5]{dB}$ net coding gain at BER of $10^{-4}$ when compared to the construction proposed in \cite{constructTalVardy}, while the only difference between the two codes is the selection of the frozen/non-frozen bit positions, i.e., both use exactly the same decoder. 

The $d_{min}$ of the \ac{BP}-tailored polar code using \ac{GenAlg} was found to be $8$, while the $d_{min}$ of the polar code constructed via \cite{constructTalVardy} is $16$.
This is due to the fact that the performance of a linear code under iterative decoding is dominated by the structure of the stopping sets in the Tanner graph of the code and not its $d_{min}$ \cite{vardyStoppingDistance}. 
Thus, $d_{min}$ is not the only parameter to be maximized in order to design linear codes tailored to iterative decoding. A fact supporting this claim is that the \ac{RM} code has a larger (maximum) $d_{min}$ but worse error-rate performance under iterative decoding when compared to polar codes \cite{ArikanBP}.

Note that the BER performance of the \ac{BP}-tailored polar code is very close to the performance of a polar code constructed via \cite{constructTalVardy} under \ac{SCL} $\left(L=32\right)$, see Fig. \ref{fig:BP_comp}.

\begin{figure}[t]
	\centering
		\includegraphics{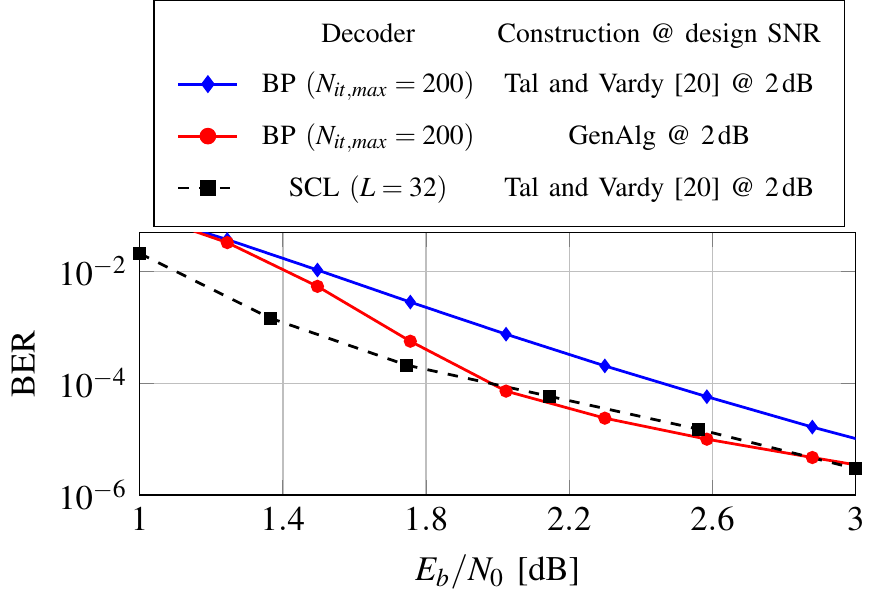}    
\vspace{-0.75cm}
\caption{\small BER of the \ac{GenAlg}-based $\mathcal{P}$(2048,1024)-code under \ac{BP} decoding over the AWGN channel.}	
\label{fig:BP_comp}
\vspace{-0.4cm}
\end{figure}
\vspace{-0.2cm}
\subsection{\ac{SCL} decoder}
Next, the \ac{GenAlg} is applied to construct polar codes tailored to \ac{SCL} $\left(L=32\right)$ over the \ac{AWGN} channel. Fig. \ref{fig:SCL_comp} shows a \ac{BER} comparison between a code constructed via the method proposed in \cite{constructTalVardy} at ${E_b}/{N_0} = \unit[2]{dB}$ and a code constructed using our proposed \ac{GenAlg} at ${E_b}/{N_0} = \unit[2]{dB}$ under \ac{SCL} $\left(L=32\right)$ decoding. The \ac{GenAlg}-based construction shows significant performance improvements, yielding a $\unit[1]{dB}$ net coding gain at BER of $10^{-6}$ when compared to the construction proposed in \cite{constructTalVardy}, where again the only difference between the two codes is the frozen/non-frozen bit positions.

The \ac{GenAlg}-optimized polar code without \ac{CRC}-aid under \ac{SCL} decoding performs equally well as the \ac{CRC}-aided polar code under CRC-aided SCL decoding, with the same list size $L=32$ and the same code rate $R_c=0.5$. 

The $d_{min}$ of the \ac{SCL}-tailored polar code using \ac{GenAlg} is $16$ which is exactly the same as the $d_{min}$ of the polar code constructed via \cite{constructTalVardy}. However, the RM-Polar code has a $d_{min}=32$.
Thus, again $d_{min}$ is not the only parameter to be maximized in order to design polar codes tailored to \ac{SCL} decoding. Note that only maximizing $d_{min}$ will lead to an \ac{RM} code.

\begin{figure}[t]
	\centering
	\resizebox{0.975\columnwidth}{!}{
		\includegraphics{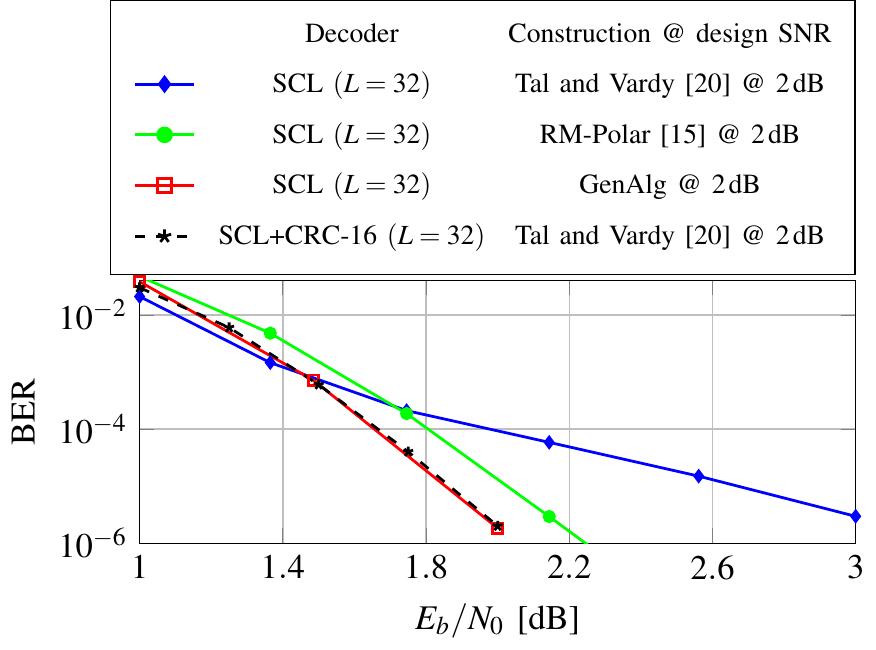}		
		}
	\vspace{-0.3cm}
	\caption{\small BER of the \ac{GenAlg}-based $\mathcal{P}$(2048,1024)-code under \ac{SCL} decoding over the \ac{AWGN} channel. The CRC-aided polar code is a $\mathcal{P}$(2048,1040)-code: $N=2048$, $k=1024$ and $r=16$.}		
	
	\label{fig:SCL_comp}
	\vspace{-0.6cm}
\end{figure}

The distribution of frozen bit-channels over the $N$ synthesized virtual channels, shown in the ``Frozen channel chart'' in Fig. \ref{fig:hist}, depends on the considered decoder while constructing the code. The conventional construction techniques (e.g., \cite{ArikanMain,constructTalVardy}) assume that an SC decoder is used. Thus, these construction techniques yield very similar $\mathbf{A}$-vectors having a very similar distribution as shown in Fig. \ref{fig:Bha-hist} and \ref{fig:TalVardy-hist}. Our proposed GenAlg-based construction tailors the $\mathbf{A}$-vector (or the code) to a specific decoder (e.g., BP or SCL). The BP decoder has an iterative (i.e., non-sequential) decoding nature when compared to SC decoding. Thus, the $\mathbf{A}$-vector tailored to BP decoding has a much different frozen bit-channel distribution as shown in Fig. \ref{fig:GenAlg-BP-hist}. Note that the $\mathbf{A}$-vector tailored to BP decoding contains a significant number of non-frozen bits in the first half part of the code, which is not the case in the  conventional code construction algorithms assuming \ac{SC} decoding. Similarly, the GenAlg-based construction tailored to SCL decoding should take the list decoding nature into consideration and, thus, a different frozen bit-channel distribution as shown in Fig. \ref{fig:GenAlg-SCL-hist}. This shows that the GenAlg-based polar code construction takes the decoding nature into consideration while constructing the code, which is the main advantage of the GenAlg.  
Furthermore, Table \ref{tab:decoder-tailored} shows that the GenAlg-optimized $\mathbf{A}$-vectors are indeed tailored to the considered decoder.
\vspace{-0.1cm}

\begin{table}[H]
	\begin{center}
		\caption{\small Illustration of polar design and decoder architecture mismatch by evaluating the minimum $E_b/N_0$ required to achieve a target BER of $10^{-4}$ for a $\mathcal{P}$(2048,1024)-code over AWGN channel}\label{tab:decoder-tailored}

		\begin{tabular}{cccc}
			\hline 
			\multirow{2}{*}{\scriptsize Construction @ design SNR} & \multicolumn{3}{c}{\scriptsize Decoder}\tabularnewline
			\cline{2-4} 
			&\scriptsize SC & \scriptsize BP $\left(N_{it,max}=200\right)$ & \scriptsize SCL $\left(L=32\right)$ \tabularnewline
			\hline 
			\scriptsize Bhattacharyya \cite{ArikanMain} @ $\unit[3.6]{dB}$ & \scriptsize \textbf{2.7 dB} & \scriptsize 2.45 dB & \scriptsize 1.8 dB \tabularnewline
			\scriptsize Tal and Vardy \cite{constructTalVardy} @ $\unit[2]{dB}$ & \scriptsize \textbf{2.65 dB} & \scriptsize 2.45 dB & \scriptsize 2 dB \tabularnewline
			\scriptsize GenAlg BP-tailored @ $\unit[2]{dB}$ & \scriptsize $>$ 9 dB & \scriptsize \textbf{2 dB} & \scriptsize $>$ 7 dB\tabularnewline
			\scriptsize GenAlg SCL-tailored @ $\unit[2]{dB}$ & \scriptsize $>$ 6 dB & \scriptsize 2.55 dB & \scriptsize \textbf{1.65 dB}\tabularnewline
			\hline 
		\end{tabular}
	\end{center}
	\vspace{-0.4cm}
\end{table}

 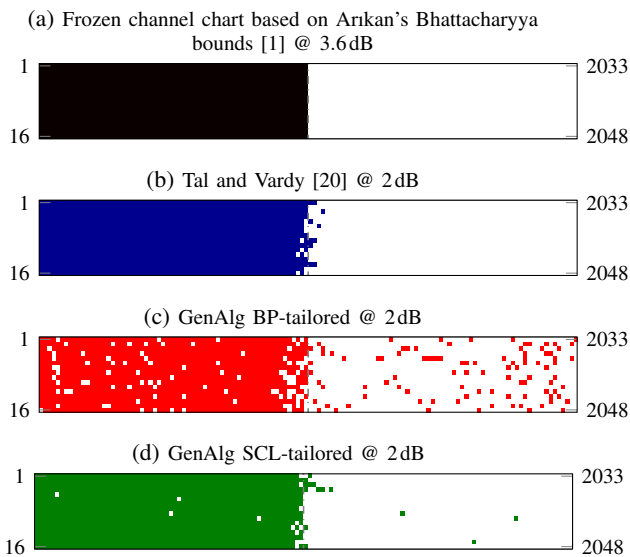
\begin{figure}[t]
 	\captionsetup[subfigure]{position=b}
 	\centering
 	
 	\begin{subfigure}{0.85\columnwidth}
 		\centering
 		\captionsetup{justification=centering}
 		\caption{\footnotesize Frozen channel chart based on Arıkan's Bhattacharyya bounds \cite{ArikanMain} @ $\unit[3.6]{dB}$} 		
 		\vspace{-0.1cm}
 		\begin{tikzpicture}
 		\begin{axis}[%
 		width=0.95\linewidth,
 		height=1cm,
 		scale only axis,
 		axis y line*=right,
 		xmin=0.5,
 		xmax=128.5,
 		yticklabels={2033,2048},ytick = {16,1},
 		xtick =\empty,
 		tick label style = {font = \footnotesize}, 		
 		ymin=0.5,
 		ymax=16.5,
 		axis background/.style={fill=white},
 		]
 		\end{axis}
 		\begin{axis}[%
 		width=0.95\linewidth,
 		height=1cm,
 		scale only axis,
 		axis on top,
 		xmin=0.5,
 		xmax=128.5,
 		xtick={\empty},
 		y dir=reverse,
 		ymin=0.5,
 		ymax=16.5,
 		ytick={ 1, 16},
 		tick label style = {font = \footnotesize}, 	 		
 		axis background/.style={fill=white},
 		]
 		\addplot [forget plot] graphics [xmin=0.5, xmax=128.5, ymin=0.5, ymax=16.5] {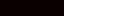};
 		\addplot +[mark=none,gray,dash dot,semithick] coordinates {(64.5, -1) (64.5, 18)};
 		\end{axis}
 		\end{tikzpicture}%
 		\label{fig:Bha-hist}
 	\end{subfigure}
 	
 	
 	\begin{subfigure}{0.85\columnwidth}  
 		\centering 
 		\caption{\footnotesize Tal and Vardy \cite{constructTalVardy} @ $\unit[2]{dB}$}   
 		\vspace{-0.1cm}
 		\begin{tikzpicture}
 		\begin{axis}[%
 		width=0.95\linewidth,
 		height=1cm,
 		scale only axis,
 		axis y line*=right,
 		xmin=0.5,
 		xmax=128.5,
 		yticklabels={2033,2048},ytick = {16,1},
 		xtick =\empty,
 		tick label style = {font = \footnotesize}, 	 		
 		ymin=0.5,
 		ymax=16.5,
 		axis background/.style={fill=white},
 		]
 		\end{axis}
 		\begin{axis}[%
 		width=0.95\linewidth,
 		height=1cm,
 		scale only axis,
 		axis on top,
 		xmin=0.5,
 		xmax=128.5,
 		xtick={\empty},
 		y dir=reverse,
 		ymin=0.5,
 		ymax=16.5,
 		ytick={ 1, 16},
 		tick label style = {font = \footnotesize}, 	 		
 		axis background/.style={fill=white},
 		title style={font=\bfseries},
 		]
 		\addplot [forget plot] graphics [xmin=0.5, xmax=128.5, ymin=0.5, ymax=16.5] {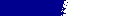};
 		\addplot +[mark=none,gray,dash dot,semithick] coordinates {(64.5, -1) (64.5, 18)};
 		\end{axis}
 		\end{tikzpicture}%
 		\label{fig:TalVardy-hist}
 	\end{subfigure}
 	
 	
 	\begin{subfigure}{0.85\columnwidth}   
 		\centering
 		\caption{\footnotesize GenAlg BP-tailored @ $\unit[2]{dB}$}  
 		\vspace{-0.1cm}
 		\begin{tikzpicture}
 		\begin{axis}[%
 		width=0.95\linewidth,
 		height=1cm,
 		scale only axis,
 		axis y line*=right,
 		xmin=0.5,
 		xmax=128.5,
 		yticklabels={2033,2048},ytick = {16,1},
 		xtick =\empty,
 		tick label style = {font = \footnotesize}, 	 		
 		ymin=0.5,
 		ymax=16.5,
 		axis background/.style={fill=white},
 		]
 		\end{axis}
 		\begin{axis}[%
 		width=0.95\linewidth,
 		height=1cm,
 		scale only axis,
 		axis on top,
 		xmin=0.5,
 		xmax=128.5,
 		xtick={\empty},
 		y dir=reverse,
 		ymin=0.5,
 		ymax=16.5,
 		ytick={ 1, 16},
 		tick label style = {font = \footnotesize}, 	 		
 		axis background/.style={fill=white},
 		title style={font=\bfseries},
 		]
 		\addplot [forget plot] graphics [xmin=0.5, xmax=128.5, ymin=0.5, ymax=16.5] {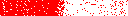};
 		\addplot +[mark=none,gray,dash dot,semithick] coordinates {(64.5, -1) (64.5, 18)};
 		\end{axis}
 		\end{tikzpicture}%
 		\label{fig:GenAlg-BP-hist}
 	\end{subfigure}
 	
 	
 	\begin{subfigure}{0.85\columnwidth}   
 		\centering 
 		\caption{\footnotesize GenAlg SCL-tailored @ $\unit[2]{dB}$}     
 		\vspace{-0.1cm}
 		\begin{tikzpicture}
 		\begin{axis}[%
 		width=0.95\linewidth,
 		height=1cm,
 		scale only axis,
 		axis y line*=right,
 		xmin=0.5,
 		xmax=128.5,
 		yticklabels={2033,2048},ytick = {16,1},
 		xtick =\empty,
 		tick label style = {font = \footnotesize}, 	 		
 		ymin=0.5,
 		ymax=16.5,
 		axis background/.style={fill=white},
 		]
 		\end{axis}
 		\begin{axis}[%
 		width=0.95\linewidth,
 		height=1cm,
 		scale only axis,
 		axis on top,
 		xmin=0.5,
 		xmax=128.5,
 		xtick={\empty},
 		y dir=reverse,
 		ymin=0.5,
 		ymax=16.5,
 		ytick={ 1, 16},
 		tick label style = {font = \footnotesize}, 	 		
 		axis background/.style={fill=white},
 		title style={font=\bfseries},
 		]
 		\addplot [forget plot] graphics [xmin=0.5, xmax=128.5, ymin=0.5, ymax=16.5] {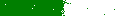};
 		\addplot +[mark=none,gray,dash dot,semithick] coordinates {(64.5, -1) (64.5, 18)};
 		\end{axis}
 		\end{tikzpicture}%
 		\label{fig:GenAlg-SCL-hist}
 	\end{subfigure}
 	\vspace{-0.2cm}
 	\caption{\small Frozen channel chart (i.e., frozen bit position pattern) of a $\mathcal{P}$(2048,1024)-code over the AWGN channel with different polar code construction algorithms. 
 		The $2048$ bit positions are plotted over a $16 \times 128$ matrix and sorted with decreasing Bhattacharyya parameter value. 
 		White: non-frozen; colored: frozen.} 
 	\label{fig:hist}
 	\vspace{-0.5cm}
 \end{figure}

\section{Conclusion} \label{sec:conc}
\vspace{-0.1cm}
As stated in \cite{talvardyList}, the best error-rate performance of finite length polar codes (i.e., under \ac{ML} decoding without \ac{CRC}) is not competitive with the other state-of-the-art coding schemes. In this work, we focus on polar code construction (i.e., changing the code itself which is defined by the $\mathbf{A}$-vector) in order to boost the error-rate performance under the state-of-the-art feasible practical decoders (e.g., \ac{BP}, \ac{SCL}). We propose a Genetic Algorithm-based polar code design where the decoding nature is taken into account, yielding significant performance gains in terms of error-rate. 
For a polar code of length $2048$ and code rate $0.5$ over the binary input \ac{AWGN} channel under \emph{plain} \ac{SCL} decoding, approximately a $\unit[1]{dB}$ coding gain at \ac{BER} of $10^{-6}$ is achieved when compared to the conventionally constructed polar codes. This enables achieving the same error-rate performance of polar codes under state-of-the-art CRC-aided SCL decoding with \emph{plain} \ac{SCL} decoding without the aid of a CRC. 
Furthermore, \ac{GenAlg} is used to construct polar codes tailored to flooding \ac{BP} decoding, finally closing the performance gap between conventional iterative \ac{BP} decoding and conventional \ac{SCL} decoding. Fortunately, the achieved gains come ``for free'', since optimizing the $\mathbf{A}$-vector is done once and offline. The source code and the best polar code designs from this work can be found in \cite{GenAlg_Github}.


\bibliographystyle{IEEEtran}
\bibliography{references}
\end{NoHyper}
\end{document}